\documentclass[10pt,twocolumn,twoside]{IEEEtran}
\usepackage{setspace}
\usepackage{amsmath}
\usepackage{algorithm}
\usepackage{graphicx}
\usepackage{color}
\usepackage{epsfig}
\usepackage{subfigure}

\title{Dictionary Optimization for Block-Sparse Representations
\thanks{The authors are with the Technion - Israel Institute of
Technology, Haifa, Israel. Email: kevin@tx.technion.ac.il,
lihi@ee.technion.ac.il, yonina@ee.technion.ac.il.}}
\author{Kevin Rosenblum, Lihi Zelnik-Manor, Yonina C. Eldar}

\begin{document}

\maketitle

\begin{abstract}
Recent work has demonstrated that using a carefully designed
dictionary instead of a predefined one, can improve the sparsity in
jointly representing a class of signals. This has motivated the
derivation of learning methods for designing a dictionary which
leads to the sparsest representation for a given set of signals. In
some applications, the signals of interest can have further
structure, so that they can be well approximated by a union of a
small number of subspaces (e.g., face recognition and motion
segmentation). This implies the existence of a dictionary which
enables {\em block-sparse} representations of the input signals once
its atoms are properly sorted into blocks. In this paper, we propose
an algorithm for learning a {\em block-sparsifying dictionary} of a
given set of signals. We do {\em not} require prior knowledge on the
association of signals into groups (subspaces). Instead, we develop
a method that automatically detects the underlying block structure.
This is achieved by iteratively alternating between updating the
block structure of the dictionary and updating the dictionary atoms
to better fit the data. Our experiments show that for block-sparse
data the proposed algorithm significantly improves the dictionary
recovery ability and lowers the representation error compared to
dictionary learning methods that do not employ block structure.
\end{abstract}

%
%

\section{Introduction}
\label{sec:intro}

The framework of sparse coding aims at recovering an unknown vector
$\theta \in R^K$ from an under-determined system of linear equations
$x = D \theta$, where $D \in R^{N \times K}$ is a dictionary, and $x
\in R^N$ is an observation vector with $N<K$. Since the system is
under-determined, $\theta$ can not be recovered without additional
information. The framework of compressed sensing \cite{CS1,CS2}
exploits sparsity of $\theta$ in order to enable recovery.
Specifically, when $\theta$ is known to be sparse so that it
contains few nonzero coefficients, and when $D$ is chosen properly,
then $\theta$ can be recovered uniquely from $x = D \theta$.
Recovery is possible irrespectively of the locations of the nonzero
entries of $\theta$. This result has given rise to a multitude of
different recovery algorithms. Most prominent among them are Basis
Pursuit (BP) \cite{BP, CS1} and Orthogonal Matching Pursuit (OMP)
\cite{OMP1,OMP2}.

Recent work  \cite{KSVD1,MOD,Dic1,Dic2,sapiro} has demonstrated that
adapting the dictionary $D$ to fit a given set of signal examples
leads to improved signal reconstruction. At the price of being slow,
these learning algorithms attempt to find a dictionary that leads to
optimal sparse representations for a certain class of signals. These
methods show impressive results for representations with arbitrary
sparsity structures. In some applications, however, the
representations have a unique sparsity structure that can be
exploited. Our interest is in the case of signals that are known to
be drawn from a union of a small number of subspaces
\cite{BBP1,kfir}. This occurs naturally, for example, in face
recognition \cite{face1,face2}, motion segmentation \cite{motion},
multiband signals \cite{mult1,mult2,mult3}, measurements of gene
expression levels \cite{DNA}, and more. For such signals, sorting
the dictionary atoms according to the underlying subspaces leads to
sparse representations which exhibit a block-sparse structure, i.e.,
the nonzero coefficients occur in clusters of varying sizes. Several
methods, such as \emph{Block BP} (BBP) \cite{BBP1,BBP2,BBP3} and
\emph{Block OMP} (BOMP) \cite{BOMP1,BOMP2} have been proposed to
take advantage of this structure in recovering the block-sparse
representation $\theta$. These methods typically assume that the
dictionary is predetermined and the block structure is known.

In this paper we propose a method for designing a
\emph{block-sparsifying dictionary} for a given set of signals. In
other words, we wish to find a dictionary that provides block-sparse
representations best suited to the signals in a given set. To take
advantage of the block structure via block-sparse approximation
methods, it is necessary to know the block structure of the
dictionary. We do not assume that it is known a-priori. Instead, we
infer the block structure from the data while adapting the
dictionary.

We start by formulating this task as an optimization problem. We
then present an algorithm for minimizing the proposed objective,
which iteratively alternates between updating the block structure
and updating the dictionary. The block structure is inferred by the
agglomerative clustering of dictionary atoms that induce similar
sparsity patterns. In other words, after finding the sparse
representations of the training signals, the atoms are progressively
merged according to the similarity of the sets of signals they
represent. A variety of segmentation methods through subspace
modeling have been proposed recently \cite{SUB1,SUB2,SUB3}. These
techniques learn an underlying collection of subspaces based on the
assumption that each of the samples lies close to one of them.
However, unlike our method, they do not treat the more general case
where the signals are drawn from a union of several subspaces.

The dictionary blocks are then sequentially updated to minimize the
representation error at each step. The proposed algorithm is an
intuitive extension of the K-SVD algorithm \cite{KSVD1}, which
yields sparsifying dictionaries by sequentially updating the
dictionary atoms, to the case of block structures. In other words,
when the blocks are of size $1$ our cost function and the algorithm
we propose reduce to K-SVD. Our experiments show that updating the
dictionary block by block is preferred over updating the atoms in
the dictionary one by one, as in K-SVD.

We show empirically that both parts of the algorithm are
indispensable to obtain high performance. While fixing a random
block structure and applying only the dictionary update part leads
to improved signal reconstruction compared to K-SVD, combining the
two parts leads to even better results. Furthermore, our experiments
show that K-SVD often fails to recover the underlying block
structure. This is in contrast to our algorithm which succeeds in
detecting most of the blocks.

We begin by reviewing previous work on dictionary design in
Section~\ref{sec:prior-work}. In Section~\ref{ssec:problem} we
present an objective for designing block-sparsifying dictionaries.
We show that this objective is a direct extension of the one used by
K-SVD. We then propose an algorithm for minimizing the proposed cost
function (Section~\ref{ssec:alg}). In Section~\ref{ssec:SAC} we give
a detailed description of the algorithm for finding a block
structure and in Section~\ref{ssec:BKSVD} we describe the dictionary
update part. We evaluate the performance of the proposed algorithms
and compare them to previous work in Section~\ref{sec:experiments}.

Throughout the paper, we denote vectors by lowercase letters, e.g.,
$x$, and matrices by uppercase letters, e.g., $A$. The $j$th column
of a matrix $A$ is written as $A_j$, and the $i$th row as $A^i$. The
sub-matrix containing the entries of $A$ in the rows with indices
$r$ and the columns with indices $c$ is denoted $A^r_c$. The
Frobenius norm is defined by
$\|A\|_F\equiv\sqrt{\sum_j{\|A_j\|_2^2}}$. The $i$th element of a
vector $x$ is denoted $x[i]$. $\|x\|_p$ is its $l_p$-norm and
$\|x\|_0$ counts the number of non-zero entries in $x$.

%
%

\section{Prior work on dictionary design}
\label{sec:prior-work}

The goal in dictionary learning is to find a dictionary $D$ and a
representation matrix $\Theta$ that best match a given set of
vectors $X_i$ that are the columns of $X$. In addition, we would
like each vector $\Theta_i$ of $\Theta$ to be sparse. In this
section we briefly review two popular sparsifying dictionary design
algorithms, K-SVD \cite{KSVD1,KSVD2} and MOD (Method of Optimal
Directions) \cite{MOD}. We will generalize these methods to
block-sparsifying dictionary design in Section~\ref{sec:dic}.

To learn an optimal dictionary, both MOD and K-SVD attempt to
optimize the same cost function for a given sparsity measure $k$:
\begin{eqnarray}
\min_{D,\Theta}&&\|X-D\Theta\|_F \nonumber\\
\textrm{s.t.}&&\|\Theta_i\|_0\leq k, \textrm{\ } i=1,\ldots,L
\label{Dicopt}
\end{eqnarray}
where $X \in R^{N \times L}$ is a matrix containing $L$ given input
signals, $D \in R^{N \times  K}$ is the dictionary and $\Theta \in
R^{K \times L}$ is a sparse representation of the signals. Note that
the solution of \eqref{Dicopt} is never unique due to the invariance
of $D$ to permutation and scaling of columns. This is partially
resolved by requiring normalized columns in $D$. We will therefore
assume throughout the paper that the columns of $D$ are normalized
to have $l_2$-norm equal $1$.

Problem \eqref{Dicopt} is non-convex and NP-hard in general. Both
MOD and K-SVD attempt to approximate \eqref{Dicopt} using a
relaxation technique which iteratively fixes all the parameters but
one, and optimizes the objective over the remaining variable. In
this approach the objective decreases (or is left unchanged) at each
step, so that convergence to a local minimum is guaranteed. Since
this might not be the global optimum both approaches are strongly
dependent on the initial dictionary $D^{(0)}$. The convention is to
initialize $D^{(0)}$ as a collection of $K$ data signals from the
same class as the training signals $X$.

The first step of the $n$th iteration in both algorithms optimizes
$\Theta$ given a fixed dictionary $D^{(n-1)}$, so that
\eqref{Dicopt} becomes:
\begin{eqnarray}
\Theta^{(n)}=\arg\min_{\Theta}&&\|X-D^{(n-1)}\Theta\|_F\nonumber\\
\textrm{s.t.}&&\|\Theta_i\|_0\leq k, \textrm{\ } i=1,\ldots,L.
\end{eqnarray}
This problem can be solved approximately using sparse coding methods
such as BP or OMP for each column of $\Theta$, since the problem is
separable in these columns. Next, $\Theta^{(n)}$ is kept fixed and
the representation error is minimized over $D$:
\begin{equation}
D^{(n)}=\arg\min_D\|X-D\Theta^{(n)}\|_F. \label{Dic_step}
\end{equation}
The difference between MOD and K-SVD lies in the choice of
optimization method for $D^{(n)}$. While K-SVD converges faster than
MOD, both methods yield similar results.

The MOD algorithm treats the problem in  \eqref{Dic_step} directly.
This problem has a closed form solution given by the pseudo-inverse:
\begin{equation}
D^{(n)}=X\Theta'^{(n)}(\Theta^{(n)}\Theta'^{(n)})^{-1}.
\end{equation}
Here we assume for simplicity that $\Theta^{(n)}\Theta'^{(n)}$ is
invertible. The K-SVD method solves \eqref{Dic_step} differently.
The columns in $D^{(n-1)}$ are updated sequentially, along with the
corresponding non-zero coefficients in $\Theta^{(n)}$. This parallel
update leads to a significant speedup while preserving the sparsity
pattern of $\Theta^{(n)}$. For $j=1,\ldots,K$, the update is as
follows. Let $\omega_j \equiv \{i \in 1,\ldots,L|\Theta^j_i \neq
0\}$ be the set of indices corresponding to columns in
$\Theta^{(n)}$ that use the atom $D_j$, i.e., their $i$th row is
non-zero. Denote by $R_{\omega_j}=X_{\omega_j}-\sum_{i \neq
j}(D_i\Theta^i_{\omega_j})$ the representation error of the signals
$X_{\omega_j}$ excluding the contribution of the $j$th atom. The
representation error of the signals with indices $\omega_j$ can then
be written as $\|R_{\omega_j}-D_j\Theta^j_{\omega_j}\|_F$. The goal
of the update step is to minimize this representation error, which
is accomplished by choosing
\[D_j=U_1, \textrm{\ \ \ \ } \Theta^j_{\omega_j}=\Delta_1^1V_1'.\]
Here $U\Delta V'$ is the Singular Value Decomposition (SVD) of
$R_{\omega_j}$. Note, that the columns of $D$ remain normalized
after the update. The K-SVD algorithm obtains the dictionary update
by $K$ separate SVD computations, which explains its name.

\section{Block-Sparsifying Dictionary optimization}
\label{sec:dic}

We now formulate the problem of block-sparsifying dictionary design.
We then propose an algorithm which can be seen as a natural
extension of K-SVD for the case of signals with block sparse
representations. Our method involves an additional clustering step
in order to determine the block structure.

\subsection{Problem definition}
\label{ssec:problem}

For a given set of $L$ signals $X=\{X_i\}_{i=1}^L \in R^N$, we wish
to find a dictionary $D \in R^{N\times K}$ whose atoms are sorted in
blocks, and which provides the most accurate representation vectors
whose non-zero values are concentrated in a fixed number of blocks.
In previous works dealing with the block-sparse model, it is
typically assumed that the block structure in $D$ is known a-priori,
and even more specifically, that the atoms in $D$ are sorted
according to blocks \cite{BBP1,BBP2}. Instead, in this paper we
address the more general case where the block structure is unknown
and the blocks can be of varying sizes. The only assumption we make
on the block structure is that the maximal block size, denoted by
$s$, is known.

\begin{figure}
\centering
\includegraphics[width=100mm,angle=0,scale=0.8]{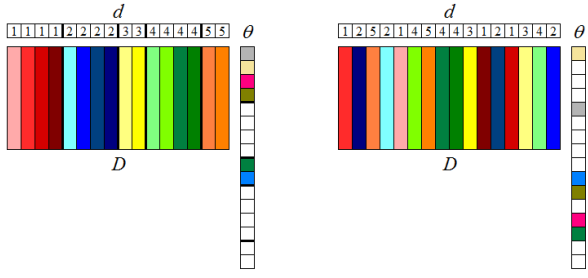}
\caption{Two equivalent examples of dictionaries $D$ and block
structures $d$ with $5$ blocks, together with $2$-block-sparse
representations $\theta$. Both examples represent the same signal,
since the atoms in $D$ and the entries of $d$ and $\theta$ are
permuted in the same manner.} \label{fig:block}
\end{figure}

More specifically, suppose we have a dictionary whose atoms are
sorted in blocks that enable {\em block-sparse} representations of
the input signals. Assume that each block is given an index number.
Let $d \in R^K$ be the vector of block assignments for the atoms of
$D$, i.e., $d[i]$ is the block index of the atom $D_i$. We say that
a vector $\theta \in R^K$ is $k$-block-sparse over $d$ if its
non-zero values are concentrated in $k$ blocks only. This is denoted
by $\|\theta\|_{0,d}=k$, where $\|\theta\|_{0,d}$ is the $l_0$-norm
over $d$ and counts the number of non-zero blocks as defined by $d$.
Fig. \ref{fig:block} presents examples of two different block
structures and two corresponding block-sparse vectors and
dictionaries.

Our goal is to find a dictionary $D$ and a block structure $d$, with
maximal block size $s$, that lead to optimal $k$-block sparse
representations $\Theta=\{\Theta_i\}_{i=1}^L$ for the signals in
$X$:
\begin{eqnarray}
\min_{D,d,\Theta}&&\|X-D\Theta\|_F \nonumber\\
\textrm{s.t.}&& \|\Theta_i\|_{0,d}\leq k, \textrm{\ } i=1,\ldots,L \nonumber\\
&& |d_j| \leq s, \textrm{\ } j \in d \label{blockDicopt}
\end{eqnarray}
where $d_j=\{i\in 1,\ldots,K|d[i]=j\}$ is the set of indices
belonging to block $j$ (i.e., the list of atoms in block $j$).

The case when there is no underlying block structure or when the
block structure is ignored, is equivalent to setting $s=1$ and
$d=[1,\ldots,K]$. Substituting this into \eqref{blockDicopt},
reduces it to \eqref{Dicopt}. In this setting, the objective and the
algorithm we propose coincide with K-SVD. In
Section~\ref{sec:experiments} we demonstrate through simulations
that when an underlying block structure exists, optimizing
\eqref{blockDicopt} via the proposed framework improves recovery
results and lowers the representation errors with respect to
\eqref{Dicopt}.

\subsection{Algorithm Preview}
\label{ssec:alg}

In this section, we propose a framework for solving
\eqref{blockDicopt}. Since this optimization problem is non-convex,
we adopt the coordinate relaxation technique. We initialize the
dictionary $D^{(0)}$ as the outcome of the K-SVD algorithm (using a
random collection of $K$ signals leads to similar results, but
slightly slower convergence). Then, at each iteration $n$ we perform
the following two steps:

\begin{enumerate}

\item Recover the block structure by solving \eqref{blockDicopt} for $d$ and $\Theta$ while keeping $D^{(n-1)}$ fixed:
  \begin{eqnarray}
    \label{block_structure}
    [d^{(n)},\Theta^{(n)}] = \min_{d,\Theta}&&\|X-D^{(n-1)}\Theta\|_F \\
    \textrm{s.t.}&&\|\Theta_i\|_{0,d}\leq k, \textrm{\ } i=1,\ldots,L \nonumber\\
    &&|d_j| \leq s, \textrm{\ } j \in d.\nonumber
  \end{eqnarray}
An exact solution would require a combinatorial search over all
feasible $d$ and $\Theta$. Instead, we propose a tractable
approximation to \eqref{block_structure} in Section \ref{ssec:SAC},
referred to as {\em Sparse  Agglomerative Clustering (SAC)}.
Agglomerative clustering builds blocks by progressively merging the
closest atoms according to some distance metric \cite{duda, agg}.
SAC uses the $l_0$-norm for this purpose.

\item Fit the dictionary $D^{(n)}$ to the data by solving \eqref{blockDicopt} for $D$ and $\Theta$ while keeping $d^{(n)}$ fixed:
\begin{eqnarray}
    [D^{(n)},\Theta^{(n)}]=\min_{D,\Theta} &&\|X-D\Theta\|_F \\
    \textrm{s.t.} &&\|\Theta_i\|_{0,d^{(n)}}\leq k, \textrm{\ } i=1,\ldots,L\nonumber.
    \label{BKSVDobj}
  \end{eqnarray}
In Section \ref{ssec:BKSVD} we propose an algorithm, referred to as
{\em Block K-SVD (BK-SVD)}, for solving \eqref{BKSVDobj}. This
technique can be viewed as a generalization of K-SVD since the
blocks in $D^{(n)}$ are sequentially updated together with the
corresponding non-zero blocks in $\Theta^{(n)}$.
\end{enumerate}

In the following sections we describe in detail the steps of this
algorithm. The overall framework is summarized in Algorithm 1.

\begin{algorithm}
\caption{Block-Sparse Dictionary Design} \emph{Input}: A set of
signals $X$, block sparsity $k$ and maximal block size $s$.\\
\emph{Task}: Find a dictionary $D$, block structure $d$ and the
 corresponding sparse representation $\Theta$ by optimizing:
\begin{eqnarray*}
\min_{D,d,\Theta} &&\|X-D\Theta\|_F \\
\textrm{s.t.}&& \|\Theta_i\|_{0,d}\leq k, \textrm{\ } i=1,\ldots,L\\
&&|d_j| \leq s, \textrm{\ } j\in d.
\end{eqnarray*}
\emph{Initialization}: Set the initial dictionary $D^{(0)}$ as the
outcome of K-SVD.
\\\emph{Repeat from $n=1$ until convergence}:
\begin{enumerate}
  \item Fix $D^{(n-1)}$, and update $d^{(n)}$ and $\Theta^{(n)}$ by applying Sparse Agglomerative Clustering.
  \item Fix $d^{(n)}$, and update $D^{(n)}$ and $\Theta^{(n)}$ by applying BK-SVD.
  \item $n=n+1$.
\end{enumerate}
\end{algorithm}

\subsection{Block Structure Recovery: Sparse Agglomerative Clustering}
\label{ssec:SAC}

\begin{figure}
\centering \subfigure[]{
\includegraphics[width=125mm,angle=0,scale=0.5]{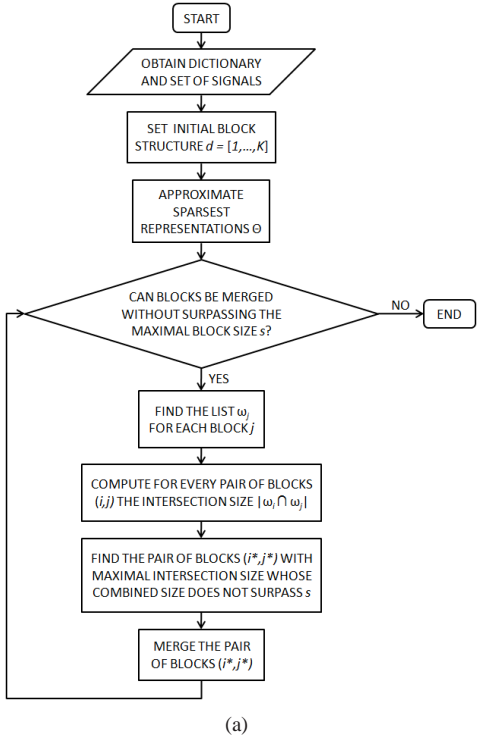}
\label{fig:SACflow} } \hspace{1cm} \subfigure[]{
\includegraphics[width=125mm,angle=0,scale=0.5]{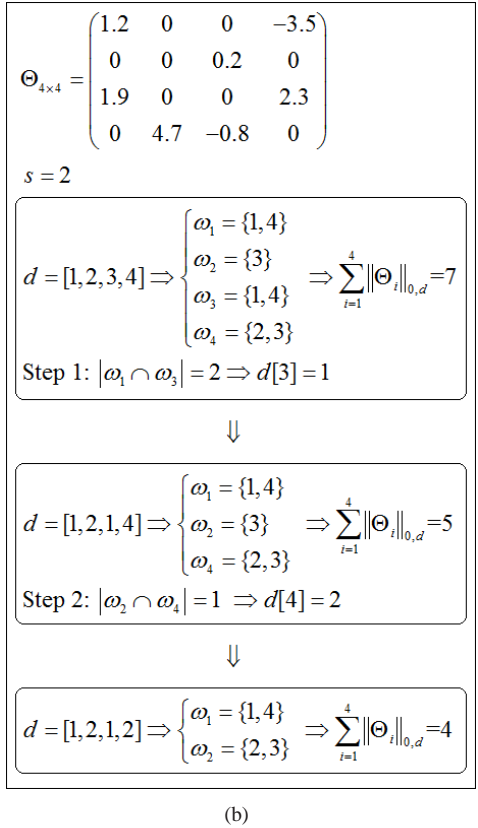}
\label{fig:SACexample} } \caption{(a) A flow chart describing the
SAC algorithm. (b) A detailed example of the decision making process
in the SAC algorithm.}
\end{figure}

In this section we propose a method for recovering the block
structure $d$ given a fixed dictionary $D$, as outlined in
Fig.~\ref{fig:SACflow}. The suggested method is based on the
coordinate relaxation technique to solve \eqref{block_structure}
efficiently. We start by initializing $d$ and $\Theta$. Since we
have no prior knowledge on $d$ it is initialized as $K$ blocks of
size $1$, i.e. $d=[1,\ldots,K]$. To initialize $\Theta$ we keep $d$
fixed and solve \eqref{block_structure} over $\Theta$ using OMP with
$k\times s$ instead of $k$ non-zero entries, since the signals are
known to be combinations of $k$ blocks of size $s$. Based on the
obtained $\Theta$, we first update $d$ as described below and then
again $\Theta$ using BOMP \cite{BOMP1}. The BOMP algorithm
sequentially selects the dictionary blocks that best match the input
signals $X_i$, and can be seen as a generalization of the OMP
algorithm to the case of blocks.

To update $d$ we wish to solve \eqref{block_structure} while keeping
$\Theta$ fixed. Although the objective does not depend on $d$, the
constraints do. Therefore, the problem becomes finding a block
structure with maximal block size $s$ that meets the constraint on
the block-sparsity of $\Theta$. To this end, we seek to minimize the
block-sparsity of $\Theta$ over $d$:
\begin{equation}
\min_d\sum_{i=1}^L\|\Theta_i\|_{0,d}
    \textrm{ \ s.t. \ }|d_j| \leq s, \textrm{\ } j \in d.
\label{block_sp_min}
\end{equation}
Before we describe how \eqref{block_sp_min} is optimized we first
wish to provide some insight. When a signal $X_i$ is well
represented by the unknown block $d_j$, then the corresponding rows
in $\Theta_i$ are likely to be non-zero. Therefore, {\em rows} of
$\Theta$ that exhibit a similar pattern of non-zeros are likely to
correspond to {\em columns} of the same dictionary block. Therefore,
grouping dictionary columns into blocks is equivalent to grouping
rows of $\Theta$ according to their sparsity pattern. To detect rows
with similar sparsity patterns we next rewrite the objective of
\eqref{block_sp_min} as a function of the pattern on non-zeros.

Let $\omega_j(\Theta,d)$ denote the list of columns in $\Theta$ that
have non-zero values in rows corresponding to block $d_j$, i.e.,
$\omega_j(\Theta,d) = \{i\in 1,\ldots,L| \textrm{ }
\|\Theta_i^{d_j}\|_2>0\}$. Problem \eqref{block_sp_min} can now be
rewritten as:
\begin{equation}
 \min_{d}\sum_{j \in d}{|\omega_j(\Theta,d)|} \textrm{ \ s.t. \ }|d_j| \leq s, \textrm{\ } j \in d
\end{equation}
where $|\omega_j|$ denotes the size of the list $\omega_j$. We
propose using a sub-optimal tractable agglomerative clustering
algorithm \cite{agg} to minimize this objective. At each step we
merge the pair of blocks that have the most similar pattern of
non-zeros in $\Theta$, leading to the steepest descent in the
objective. We allow merging blocks as long as the maximum block size
$s$ is not exceeded.

More specifically, at each step we find the pair of blocks
$(j_1^*,j_2^*)$ such that:
\begin{equation*}
[j_1^*,j_2^*] = \arg \max_{j_1 \neq j_2}
|\omega_{j_1}\cap\omega_{j_2}| \textrm{ \ s.t. \ }
|d_{j_1}|+|d_{j_2}| \leq s.
\end{equation*}
We then merge $j_1^*$ and $j_2^*$ by setting $\forall i \in d_{j_2}:
d[i]\leftarrow j_1$,
$\omega_{j_1}\leftarrow\{\omega_{j_1}\cup\omega_{j_2}\},$ and
$\omega_{j_2}\leftarrow{\o}$. This is repeated until no blocks can
be merged without breaking the constraint on the block size. We do
not limit the intersection size for merging blocks from below, since
merging is always beneficial. Merging blocks that have nothing in
common may not reduce the objective of \eqref{block_sp_min};
however, this can still lower the representation error at the next
BK-SVD iteration. Indeed, while the number of blocks $k$ stays
fixed, the number of atoms that can be used to reduce the error
increases.

Fig.~\ref{fig:SACexample} presents an example that illustrates the
notation and the steps of the algorithm. In this example the maximal
block size is $s=2$. At initialization the block structure is set to
$d=[1,2,3,4]$, which implies that the objective of
\eqref{block_sp_min} is $\sum_{i=1}^L\|\Theta_i\|_{0,d}=2+1+2+2=7$.
At the first iteration, $\omega_1$ and $\omega_3$ have the largest
intersection. Consequently, blocks $1$ and $3$ are merged. At the
second iteration, $\omega_2$ and $\omega_4$ have the largest
intersection, so that blocks $2$ and $4$ are merged. This results in
the block structure $d=[1, 2, 1, 2]$ where no blocks can be merged
without surpassing the maximal block size. The objective of
\eqref{block_sp_min} is reduced to
$\sum_{i=1}^L\|\Theta_i\|_{0,d}=4$, since all $4$ columns in
$\Theta$ are $1$-block-sparse. Note that since every column contains
non-zero values, this is the global minimum and therefore the
algorithm succeeded in solving \eqref{block_sp_min}.

While more time-efficient clustering methods exist, we have selected
agglomerative clustering because it provides a simple and intuitive
solution to our problem. Partitional clustering methods, such as
K-Means, require initialization and are therefore not suited for
highly sparse data and the $l_0$-norm metric. Moreover, since
oversized blocks are unwanted, it is preferable to limit the block
size rather than the number of blocks. It is important to note that
due to the iterative nature of our dictionary design algorithm,
clustering errors can be corrected in the following iteration, after
the dictionary has been refined.

\subsection{Block K-SVD Algorithm}
\label{ssec:BKSVD}

We now propose the BK-SVD algorithm for recovering the dictionary
$D$ and the representations $\Theta$ by optimizing \eqref{BKSVDobj}
given a block structure $d$ and input signals $X$.

Using the coordinate relaxation technique, we solve this problem by
minimizing the objective based on alternating $\Theta$ and $D$. At
each iteration $m$, we first fix $D^{(m-1)}$ and use BOMP to solve
\eqref{BKSVDobj} which reduces to
\begin{eqnarray}
\Theta^{(m)}=\arg\min_{\Theta}&&\|X-D^{(m-1)}\Theta\|_F \nonumber\\
\textrm{s.t.} && \|\Theta_i\|_{0,d}\leq k, \textrm{\ } i=1,\ldots,L.
\label{eq:optTheta}
\end{eqnarray}
Next, to obtain $D^{(m)}$ we fix $\Theta^{(m)},d$ and $X$, and
solve:
\begin{equation}
D^{(m)}=\arg\min_{D} \|X-D\Theta^{(m)}\|_F.
\end{equation}

Inspired by the K-SVD algorithm, the blocks in $D^{(m-1)}$ are
updated sequentially, along with the corresponding non-zero
coefficients in $\Theta^{(m)}$. For every block $j \in d$, the
update is as follows. Denote by $R_{\omega_{j}} = X_{\omega_j} -
\sum_{i\neq j}D_{d_i}\Theta_{\omega_j}^{d_i}$ the representation
error of the signals $X_{\omega_j}$ excluding the contribution of
the $j$th block. Here $\omega_j$ and $d_j$ are defined as in the
previous subsection. The representation error of the signals with
indices $\omega_j$ can then be written as $\|R_{\omega_j} -
D_{d_j}\Theta_{\omega_j}^{d_j}\|_F$. Finally, the representation
error is minimized by setting $D_{d_j}\Theta_{\omega_j}^{d_j}$ equal
to the matrix of rank $|d_j|$ that best approximates $R_{\omega_j}$.
This can obtained by the following updates:
\[D_{d_j}=[U_1,\ldots,U_{|d_j|}]\]
\[\Theta_{\omega_j}^{d_j}=[\Delta_1^1V_1,\ldots,\Delta_{|d_j|}^{|d_j|}V_{|d_j|}]'\]
where the $|d_j|$ highest rank components of $R_{\omega_j}$ are
computed using the SVD $R_{\omega_j}=U\Delta V'$. The updated
$D_{d_j}$ is now an orthonormal basis that optimally represents the
signals with indices $\omega_{j}$. Note that the representation
error is also minimized when multiplying $D_{d_j}$ on the right by
$W$ and $\Theta_{\omega_j}^{d_j}$ on the left by $W^{-1}$, where
$W\in R^{|d_j|\times|d_j|}$ is an invertible matrix. However, if we
require the dictionary blocks to be orthonormal subspaces, the
solution is unique up to the permutation of the atoms. It is also
important to note that if $|d_j|>|\omega_j|$, then
$|d_j|-|\omega_j|$ superfluous atoms in block $j$ can be discarded
without any loss of performance.

This dictionary update minimizes the representation error while
preserving the sparsity pattern of $\Theta^{(m)}$, as in the K-SVD
dictionary update step. However, the update step in the BK-SVD
algorithm converges faster thanks to the simultaneous optimization
of the atoms belonging to the same block. Our simulations show that
it leads to smaller representation errors as well. Moreover, the
dictionary update step in BK-SVD requires about $s$ times less SVD
computations, which makes the proposed algorithm significantly
faster than K-SVD.

We next present a simple example illustrating the advantage of the
BK-SVD dictionary update step, compared to the K-SVD update. Let
$D_1$ and $D_2$ be the atoms of the same block, of size 2. A
possible scenario is that $D_2=U_1$ and
$\Theta_{\omega_j}^{2}=-\Delta(1,1)V_1'$. In K-SVD, the first update
of $D$ is $D_1\leftarrow U_1$ and
$\Theta_{\omega_j}^1\leftarrow\Delta(1,1)V_1'$. In this case the
second update would leave $D_2$ and $\Theta_{\omega_j}^2$ unchanged.
As a consequence, only the highest rank component of $R_{\omega_j}$
is removed. Conversely, in the proposed BK-SVD algorithm, the atoms
$D_1$ and $D_2$ are updated simultaneously, resulting in the two
highest rank components of $R_{\omega_j}$ being removed.

\section{Experiments}
\label{sec:experiments}

In this section, we evaluate the contribution of the proposed
block-sparsifying dictionary design framework empirically. We also
examine the performance of the SAC and the BK-SVD algorithms
separately.

For each simulation, we repeat the following procedure $50$ times:
We randomly generate a dictionary $D^*$ of dimension ${30\times 60}$
with normally distributed entries and normalize its columns. The
block structure is chosen to be of the form:
\[d^*=[1,1,1,2,2,2,\ldots,20,20,20]\]
i.e. $D^*$ consists of $20$ subspaces of size $s=3$. We generate
$L=5000$ test signals $X$ of dimension $N=30$, that have $2$-block
sparse representations $\Theta^*$ with respect to $D^*$ (i.e.
$k=2$). The generating blocks are chosen randomly and independently
and the coefficients are i.i.d. uniformly distributed. White
Gaussian noise with varying SNR was added to $X$.

We perform three experiments:
\begin{enumerate}
  \item Given $D^*$ and $X$, we examine the ability of SAC to recover $d^*$.
  \item Given $d^*$ and $X$, we examine the ability of BK-SVD to recover $D^*$.
  \item We examine the ability of BK-SVD combined with SAC to recover $D^*$ and $d^*$ given only $X$.
\end{enumerate}

We use two measures to evaluate the success of the simulations based
on their outputs $D$, $d$ and $\Theta$:
\begin{itemize}
\item The normalized representation error
$e=\frac{\|X-D\Theta\|_F}{\|X\|_F}$.
\item The percentage $p$ of successfully recovered blocks.
For every block in $D$, we match the closest block in $D^*$ without
repetition, where the (normalized) distance between two blocks $S_1$
and $S_2$ (of sizes $s_1$ and $s_2$) is measured by:
\[\textrm{Dist}(S_1,S_2) \equiv \sqrt{\left(1-\frac{\|S_1'S_2\|_F^2}{\max(s_1,s_2)}\right)}\]
assuming that both blocks are orthonormalized. If the distance
between the block in $D$ and its matched block in $D^*$ is smaller
than $0.01$, we consider the recovery of this block as successful.
\end{itemize}

\subsection{Evaluating SAC}

\begin{figure}
\centering
\includegraphics[width=115mm,angle=0,scale=0.7]{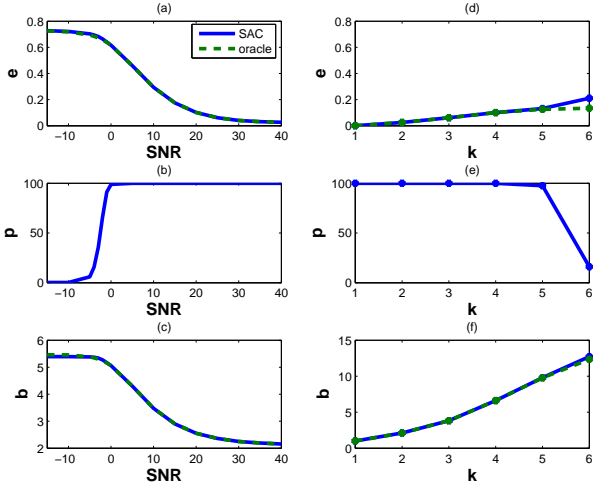}
\caption{Simulation results of the SAC algorithm. The graphs show
$e$, $p$ and $b$ as a function of the SNR of the data signals for
$k=2$ (a, b, c), and as a function of $k$ in a noiseless setting (d,
e, f).} \label{fig:SAC}
\end{figure}

To evaluate the performance of the SAC algorithm, we assume that
$D^*$ is known, and use SAC to reconstruct $d^*$ and then BOMP to
approximate $\Theta^*$. The SAC algorithm is evaluated as a function
of the SNR of the signals $X$ for $k=2$, and as a function of $k$ in
a noiseless setting. In addition to $e$ and $p$, Fig. \ref{fig:SAC}
also shows the objective of \eqref{block_sp_min}, which we denote by
$b$. We compare our results with those of an ``oracle'' algorithm,
which is given as input the true block structure $d^*$. It then uses
BOMP to find $\Theta$. The oracle's results provide a lower bound on
the reconstruction error of our algorithm (we cannot expect our
algorithm to outperform the oracle). It can be seen that for SNR
higher than $-5$[dB], the percentage $p$ of successfully recovered
blocks quickly increases to $100\%$ (Fig. \ref{fig:SAC}.(b)), the
representation error $e$ drops to zero (Fig. \ref{fig:SAC}.(a)) and
the block-sparsity $b$ drops to the lowest possible value $k=2$
(Fig. \ref{fig:SAC}.(c)). Fig.~\ref{fig:SAC}.(e) shows that the
block structure $d^*$ is perfectly recovered for $k<6$. However, for
$k=6$, SAC fails in reconstructing the block structure $d^*$, even
though the block sparsity $b$ reaches the lowest possible value
(Fig. \ref{fig:SAC}.(f)). This is a consequence of the inability of
OMP to recover the sparsest approximation of the signals $X$ with
$k\times s = 12$ nonzero entries. In terms of $e$ and $b$, our
algorithm performs nearly as good as the oracle.

\subsection{Evaluating BK-SVD}

\begin{figure}
\centering
\includegraphics[width=115mm,angle=0,scale=0.7]{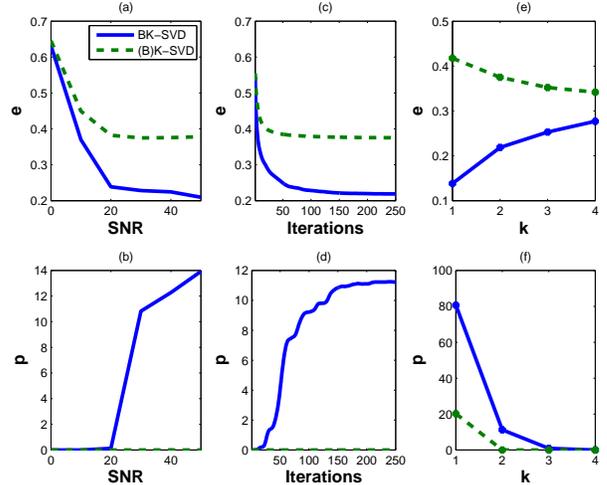}
\caption{Simulation results of the BK-SVD and (B)K-SVD algorithms.
The graphs show the reconstruction error $e$ and the recovery
percentage $p$ as a function of the SNR of the data signals for
$k=2$ and after $250$ iterations (a, b), as a function of the number
of iterations for $k=2$ in a noiseless setting (c, d), and as a
function of $k$ in a noiseless setting after $250$ iterations (e,
f).} \label{fig:BKSVD}
\end{figure}

To evaluate the performance of the BK-SVD algorithm we assume that
the block structure $d^*$ is known. We initialize the dictionary
$D^{(0)}$ by generating $20$ blocks of size $3$ where each block is
a randomly generated linear combination of $2$ randomly selected
blocks of $D^*$. We then evaluate the contribution of the proposed
BK-SVD algorithm. Recall that dictionary design consists of
iterations between two steps, updating $\Theta$ using block-sparse
approximation and updating the blocks in $D$ and their corresponding
non-zero representation coefficients. To evaluate the contribution
of the latter step, we compare its performance with that of applying
the same scheme, but using the K-SVD dictionary update step. We
refer to this algorithm as (B)K-SVD. The algorithms are evaluated as
a function of the SNR of the signals $X$ for $k=2$ after $250$
iterations, as a function of the number of iterations for $k=2$ in a
noiseless setting, and as a function of $k$ in a noiseless setting
after $250$ iterations. It is clear from Fig.~\ref{fig:BKSVD} that
the simultaneous update of the atoms in the blocks of $D$ is
imperative and does not only serve as a speedup of the algorithm.

\subsection{Evaluating the overall framework}

\begin{figure}
\centering
\includegraphics[width=115mm,angle=0,scale=0.7]{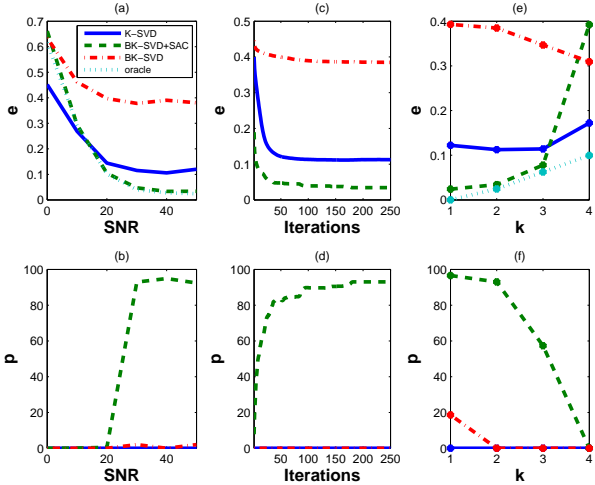}
\caption{Simulation results of our overall algorithm (BK-SVD+SAC),
the BK-SVD algorithm and the K-SVD algorithm. The graphs show the
reconstruction error $e$ and the recovery percentage $p$ as a
function of the SNR of the data signals for $k=2$ after $250$
iterations (a, b), as a function of the number of iterations for
$k=2$ in a noiseless setting (c, d), and as a function of $k$ in a
noiseless setting after $250$ iterations (e, f).}
\label{fig:BKSVDSAC}
\end{figure}

To evaluate the performance of the overall block-sparsifying
dictionary design method, we combine SAC and BK-SVD. At each
iteration we only run BK-SVD once instead of waiting for it to
converge, improving the ability of the SAC algorithm to avoid traps.
Our results are compared with those of K-SVD (with a fixed number of
$8$ coefficients) and with those of BK-SVD (with a fixed block
structure) as a function of the SNR, as a function of the number of
iterations. The algorithms are evaluated as a function of the SNR of
the signals $X$ for $k=2$ after $250$ iterations, as a function of
the number of iterations for $k=2$ in a noiseless setting, and as a
function of $k$ in a noiseless setting after $250$ iterations
(Fig.~\ref{fig:BKSVDSAC}).

Our experiments show that for SNR $>10$[dB], the proposed
block-sparsifying dictionary design algorithm yields lower
reconstruction errors (see Fig.~\ref{fig:BKSVDSAC}.(a)) and a higher
percentage of correctly reconstructed blocks (see
Fig.~\ref{fig:BKSVDSAC}.(b)), compared to K-SVD. Moreover, even in a
noiseless setting, the K-SVD algorithm fails to recover the
sparsifying dictionary, while our algorithm succeeds in recovering
$93\%$ of the dictionary blocks, as shown in
Fig.~\ref{fig:BKSVDSAC}.(d).

For SNR $\leq10$[dB] we observe that K-SVD reaches lower
reconstruction error compared to our block-sparsifying dictionary
design algorithm. This is since when the SNR is low the block
structure is no longer present in the data and the use of
block-sparse approximation algorithms is unjustified. To verify this
is indeed the cause for the failure of our algorithm, we further
compare our results with those of an oracle algorithm, which is
given as input the true dictionary $D^*$ and block structure $d^*$.
It then uses BOMP to find $\Theta$. Fig.~\ref{fig:BKSVDSAC} shows
that for all noise levels, our algorithm performs nearly as good as
the oracle. Furthermore, for SNR $\leq10$[dB] we observe that K-SVD
outperforms the oracle, implying that the use of block-sparsifying
dictionaries is unjustified. For $k<=3$, in a noiseless setting, the
performance of our algorithm lies close to that of the oracle, and
outperforms the K-SVD algorithm. However, we note that this is not
the case for $k>=4$.

Finally, we wish to evaluate the contribution of the SAC algorithm
to the overall framework. One could possibly fix an initial block
structure and then iteratively update the dictionary using BK-SVD,
in hope that this will recover the block structure.
Fig.~\ref{fig:BKSVDSAC} shows that the representation error $e$ is
much lower when including SAC in the overall framework. Moreover,
BK-SVD consistently fails in recovering the dictionary blocks.

\subsection{Choosing the maximal block size}

We now consider the problem of setting the maximal block size in the
dictionary, when all we are given is that the sizes of the blocks
are in the range $[s_l \textrm{\ } s_h]$. This also includes the
case of varying block sizes. Choosing the maximal block size $s$ to
be equal to $s_l$ will not allow to successfully reconstruct blocks
containing more than $s_l$ atoms. On the other hand, setting $s=s_h$
will cause the initial sparse representation matrix $\Theta$,
obtained by the OMP algorithm, to contain too many non-zero
coefficients. This is experienced as noise by the SAC algorithm, and
may prevent it from functioning properly. It is therefore favorable
to use OMP with $k\times s_l$ non-zero entries only, and setting the
maximal block size $s$ to be $s_h$.

\begin{figure}
\centering \subfigure[]{
\includegraphics[width=115mm,angle=0,scale=0.6]{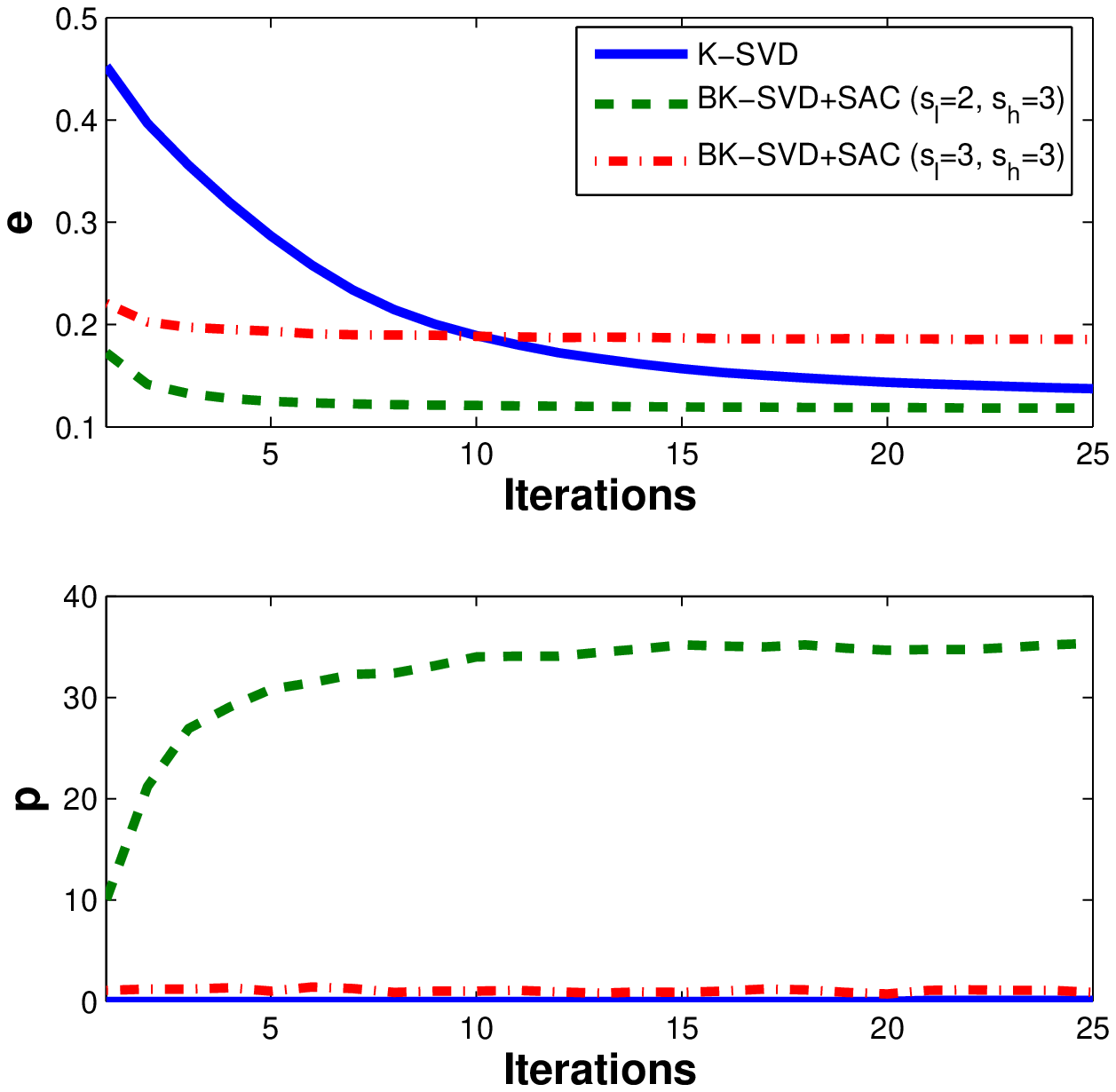}
\label{fig:mixed} } \hspace{0.2cm} \subfigure[]{
\includegraphics[width=115mm,angle=0,scale=0.6]{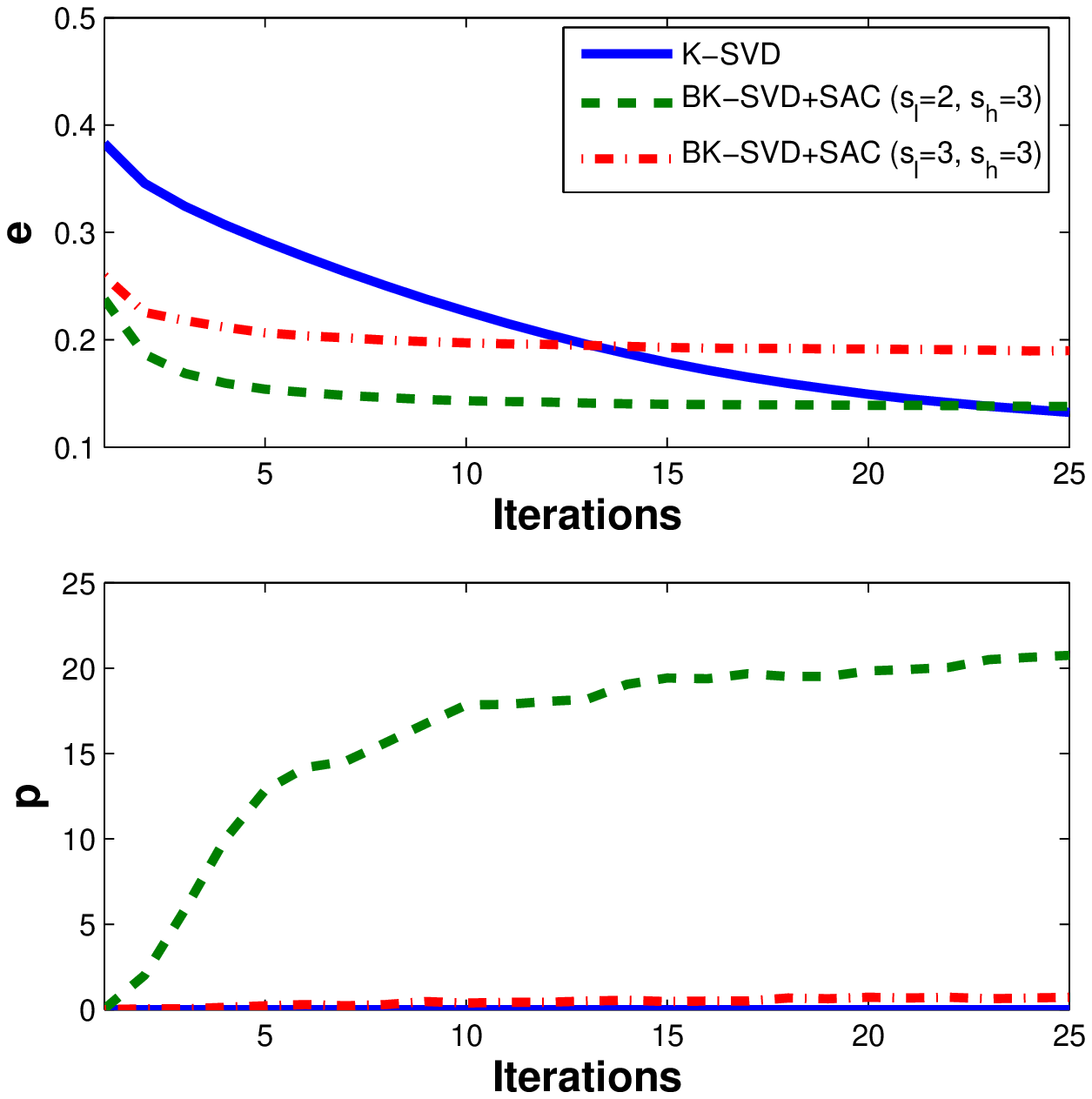}
\label{fig:wrong} } \caption{Simulation results of our overall
algorithm (BK-SVD+SAC) and the K-SVD algorithm, with maximal block
size $s_h=3$. The graphs show the reconstruction error $e$ and the
recovery percentage $p$ as a function of the number of iterations.
(a) contains $12$ blocks of size $2$ and $12$ block of size $3$. (b)
contains $30$ blocks of size $2$.}
\end{figure}

In Fig.~\ref{fig:mixed}, we evaluate the ability of our block
sparsifying dictionary design algorithm to recover the optimal
dictionary, which contains $12$ blocks of size $3$, and $12$ blocks
of size $2$. As expected, better results are obtained when choosing
$s_l=2$. In Fig.~\ref{fig:wrong}, the underlying block subspaces are
all of dimension $2$, but $s_h$ is erroneously set to be $3$. We see
that when $s_l=2$, we succeed in recovering a considerable part of
the blocks, even though blocks of size $3$ are allowed. In both
simulations, K-SVD uses $k\times s_h$ non-zero entries, which
explains why it is not significantly outperformed by our algorithm
in terms of representation error. Moreover, the percentage of
reconstructed blocks by our algorithm is relatively low compared to
the previous simulations, due to the small block sizes.

\section{Conclusions}
\label{sec:conclusions}

In this paper, we proposed a framework for the design of a
block-sparsifying dictionary given a set of signals and a maximal
block size. The algorithm consists of two steps: a block structure
update step (SAC) and a dictionary update step (BK-SVD). When the
maximal block size is chosen to be $1$, the algorithm reduces to
K-SVD.

We have shown via experiments that the block structure update step
(SAC) provides a significant contribution to the dictionary recovery
results. We have further shown that for $s>1$ the BK-SVD dictionary
update step is superior to the K-SVD dictionary update. Moreover,
the representation error obtained by our dictionary design method
lies very close to the lower bound (the oracle) for all noise
levels. This suggests that our algorithm has reached its goal in
providing dictionaries that lead to accurate sparse representations
for a given set of signals.

To further improve the proposed approach one could try and make the
dictionary design algorithm less susceptible to local minimum traps.
Another refinement could be replacing blocks in the dictionary that
contribute little to the sparse representations (i.e. ``unpopular
blocks'') with the least represented signal elements. This is
expected to only improve reconstruction results. Finally, we may
replace the time-efficient BOMP algorithm, with other block-sparse
approximation methods. We leave these issues for future research.

\section{Acknowledgements}

The research of Lihi Zelnik-Manor is supported by Marie Curie
IRG-208529.

\newpage
\bibliographystyle{IEEEbib}
\bibliography{refs}

\end{document}